\documentclass{article}

\usepackage{arxiv}

\usepackage[utf8]{inputenc} 
\usepackage[T1]{fontenc}    
\usepackage{hyperref}       
\usepackage{url}            
\usepackage{booktabs}       
\usepackage{amsfonts}       
\usepackage{nicefrac}       
\usepackage{microtype}      
\usepackage{lipsum}
\usepackage{amsmath}
\usepackage{graphicx}
\usepackage[authoryear]{natbib}

\title{Standardized Green View Index and Quantification of Different Metrics of Urban Green Vegetation}

\author{
Yusuke Kumakoshi\thanks{Corresponding author}\\
Research Center for Advanced Science and Technology\\
The University of Tokyo\\
Tokyo, 153-8904, Japan\\
\texttt{y-kuma@g.ecc.u-tokyo.ac.jp}\\
\And
Sau Yee Chan\\
Independent engineer\\
Tokyo, Japan\\
\texttt{sauyee.c@gmail.com}\\
\And
Hideki Koizumi\\
Department of Urban Engineering\\
The University of Tokyo\\
Tokyo, 113-8656, Japan\\
\texttt{hide@cd.t.u-tokyo.ac.jp}\\
\And
Xiaojiang Li\\
Department of Geography and Urban Studies\\
Temple University\\
PA 19122, U.S.A.\\
\texttt{xiaojiang.li@temple.edu}\\
\And
Yuji Yoshimura\\
Research Center for Advanced Science and Technology\\
The University of Tokyo\\
Tokyo, 153-8904, Japan\\
\texttt{yyyoshimura@cd.t.u-tokyo.ac.jp}
}

\begin{document}
\maketitle

\begin{abstract}
Urban greenery is considered an important factor in relation to sustainable development and people's quality of life in the city. Although ways to measure urban greenery have been proposed, the characteristics of each metric have not been fully established, rendering previous researches vulnerable to changes in greenery metrics. To make estimation more robust, this study aims to (1) propose an improved indicator of greenery visibility for analytical use (standardized GVI; sGVI), and (2) quantify the relation between sGVI and other greenery metrics. Analyzing a data set for Yokohama city, Japan, it is shown that the sGVI, a weighted form of GVI aggregated to an area, mitigates the bias of densely located measurement sites. Also, by comparing sGVI and NDVI at city block level, we found that sGVI captures the presence of vegetation better in the city center, whereas NDVI is better in capturing vegetation in parks and forests. These tools provide a foundation for accessing the effect of vegetation in urban landscapes in a more robust matter, enabling comparison on any arbitrary geographical scale.
\end{abstract}


\section{Introduction}
According to the \citet{un2018revision}, the proportion of the world population living in cities is expected to increase from 55\% in 2018 to 68\% by 2050. As the city becomes more densely populated, sustainable development is more indispensable than ever, in order to tackle problems such as mitigation of climate change and enhancement of the quality of life of citizens in the city. 

Amid these challenges, urban green space is expected to provide positive effects on different aspects in the city, such as reduction of air and water pollution to some extent (\citet{wakefield2001environmental,livesley2016urban,janhall2015review}), mitigation of urban heat effect\citet{norton2015planning,sanusi2017microclimate} and various sides of public health (\citet{tzoulas2007promoting,lee2011health,jesdale2013racial,nutsford2013ecological,astell2019association}). At the same time, potential discriminatory mechanism of just increasing green space in urban planning is also advocated (\citet{wolch2014urban}), and environmental injustice is empirically observed, especially in North American cities (\citet{boone2009,li2016environmental,gerrish2018relationship}). 

Green metrics occupy an essential part in these arguments, since the method by which urban greenness is measured largely affects the association analyses. Conventionally, accessibility to green spaces registered in land use data is often employed as a representation of urban greenery, but this approach does not fully capture the people's exposure to green vegetation in that street trees are ignored and different profile of vegetation in green spaces is not considered. In this point, satellite images provide an objective estimation of presence of vegetation, using Normalized Difference Visibility Index (NDVI). However, NDVI does not represent people's perception of green vegetation due to its top-down eyesight, whereas people on the street often see vegetation in horizontal direction or canopy in elevated direction (\citet{Li2015a,larkin2019evaluating}).

Recently, in order to measure eye-level visibility of green vegetation, GVI was developed (\citet{Yang2009,Li2015a}) and has been used in various studies (\citet{Li2015,li2016environmental,villeneuve2018comparing,Lu2019}). Although this index allows to capture people's perception of greenness from a specific site on street, it has been overlooked that aggregation of such site-based GVI at area level (e.g. block, census tract, administrative boundary) is sensible to the site distribution. Spatial aggregation of the site-based GVI must therefore be studied for more robust discussion on association between urban greenness and other factors.

In order to tackle this problem, this study proposes an aggregation method called standardized GVI (sGVI). By using Voronoi tessellation, the GVI sites are weighted quasi disproportionately to the density of the sites in the area of aggregation. Also, characteristics of NDVI and sGVI are compared, since comprehension of them is crucial especially applying these metrics to analytical studies. Even though different perspective between NDVI and sGVI is pointed out based on their moderate correlation (\citet{larkin2019evaluating}), their relation at spatially aggregated level must be also examined, given that such aggregation is common in association studies focusing on urban greenery.

This article is organized in the following order: after related studies are reviewed in Section \ref{section:2}, the methodologies of green metrics are presented (Section \ref{section:3}). Application of the newly proposed green metric as well as other conventional metrics and the result is discussed in Section \ref{section:4} and Section \ref{section:5}. At last, Section \ref{section:6} concludes this article.

\section{Related works}
\label{section:2}

\subsection{Green metrics}

\begin{table}[h]
\begin{tabular}{cccc}
Index     & Author & Data source& Description \\ \hline
GVI       & \begin{tabular}[c]{@{}c@{}}\citet{Yang2009}\\ \citet{Li2015a}\end{tabular} & \begin{tabular}[c]{@{}c@{}}Colored pictures \\ Street View images at different angles \end{tabular} &    \begin{tabular}[c]{@{}c@{}}Measure the proportion of \\ green pixels in each image\end{tabular}         \\ \hline
Panoramic GVI & \citet{Chen2019} & Street View panoramas & \begin{tabular}[c]{@{}c@{}}Measure the proportion of \\ green pixels in each panorama\end{tabular}  \\ \hline
Floor GVI & \citet{yu2016view} & \begin{tabular}[c]{@{}c@{}}NDVI and 3D building\\ model from LiDAR\end{tabular}  &      \begin{tabular}[c]{@{}c@{}}Measure green patches seen from \\ building floors in 3D city model\end{tabular}        \\ \hline
NDVI      &  \citet{james2015review}  & Satellite images   &   \begin{tabular}[c]{@{}c@{}}Normalize the difference of \\ red and near infrared bands\end{tabular}          
\end{tabular}
\caption{Review of green vegetation index}
\label{table:metrics}
\end{table}

The amount of greenness in a given area is traditionally quantified by land use data with green coverage or the Normalized Differential Vegetation Index (NDVI) derived from satellite imagery and the use of infrared light (\citet{james2015review}, \citet{gascon2016normalized}). 

To account for greenness underestimated by land use data or the NDVI, such as urban forests, \citet{Yang2009} proposed the Green View Index, which makes use of colored pictures to assess street-level visibility of green vegetation. This index is further elaborated in \citet{Li2015a}, by developing an automated program to estimate the visibility of greenness using the Google Street View API. Their work has been applied to cities around the world, with the computed results showcased in a website\footnote{\href{http://senseable.mit.edu/treepedia}{http://senseable.mit.edu/treepedia}}. 

However, the accuracy of green vegetation detection using the GVI was lacking, as artificial green objects can be erroneously classified as green vegetation (\citet{Li2015a}). To improve the accuracy, the use of advanced image recognition technologies such as semantic segmentation (\citet{cai2018treepedia}) and deep convolutional neural networks (\citet{cai2019quantifying}) have been explored.

Variants of the GVI have been proposed. \citet{Chen2019} proposed the panoramic GVI (PGVI), which processes the entire panorama image at a site and calculates the proportion of green objects in terms of pixel. Apart from street-level visibility, the Floor Green View Index (\citet{yu2016view}) measures green patches seen from a building floor, using LiDAR and 3D modeling data for buildings and NDVI for vegetation. The floor GVI focuses on the visibility of greenness from a building, without considering physical interaction with vegetation.

Table \ref{table:metrics} summarizes the previously proposed indices to capture greenness, even though the scopes are different each other. 

\subsection{Application of green metrics}
GVI is often studied alongside social, economic, and physiological factors. For social aspects, the survey conducted by  \citet{villeneuve2018comparing} found that GVI was positively associated with participation in summer recreational activities. For economic aspects, \citet{li2016environmental} demonstrated the correlation between economic inequality and environmental inequality in terms of accessibility to urban green vegetation, as measured by GVI\footnote{But \citet{li2016environmental} proved that there is no significant environmental disparity among racial/ethnic groups in general, in terms of accessibility to urban green vegetation.}. For physiological aspects, GVI has been associated with physical activities (\citet{li2018associations}, \citet{Lu2019}) and geriatric depression (\citet{helbich2019using}).

When compared with other factors, GVI must be aggregated to the size of the area containing the relevant data\footnote{Note that statistic data is often spatially aggregated in order to protect individual privacy.}. This is because GVI is measured at street-level sites; if aggregated at the area level, for instance block, census tract, and other administrative boundaries, the area will have several GVI values, which may be different from each other. However, there exists no consensus on the aggregation method: some studies used the median (\citet{li2016environmental}), while others used the mean (\citet{lu2018effect}, \citet{li2018associations}). As is discussed later in this article, the estimation by aggregation is affected by the method of aggregation. It is thus important to conceive a more robust aggregation method.

Furthermore, the reason behind characteristics shown by different greenery metrics have not been thoroughly understood. A survey conducted by \citet{villeneuve2018comparing} found that GVI (and not the NDVI) was positively associated with participation in recreational activities during the summer. However, it is not clear why this was the case. \citet{ye2019measuring} compared greenness measured by NDVI and a modified version of GVI considering accessibility, and found that greenness in well-developed neighborhoods tend to be underestimated by NDVI. Understanding the different characteristics is crucial on how to apply GVI at the city level in accordance with other metrics such as NDVI.

Given these gaps, this study aims to establish a method for such area-based study that mitigates bias from spatial distribution of GVI sites. Also, we try to characterize this new metric in comparison to other existing metrics.

\section{Methodology}
\label{section:3}
This section provides definitions of green vegetation metrics that are used in this study.

\subsection{Green View Index}
The Green View Index (GVI) was originally proposed by \citet{Yang2009} and developed by \citet{Li2015a} as a metric to measure the visibility of green vegetation in landscape. In this study, following \citet{Li2015a}, we process images extracted from Google Street View (GSV) for each site, and calculates the Green View using this formula:
\begin{equation}
    \text{Green View} = \frac{\sum^{n}_{i=1} Area_{g_i}}{\sum^{n}_{i=1} Area_{t_i}}\\
\end{equation}
where $n$ is the number of images for each site, set to 6 in this study\footnote{The reason for setting n as 6 is the ability to capture all surrounding scenery of the site. In contrast, setting n as 4, namely $0^\circ$, $90^\circ$, $180^\circ$ and $270^\circ$, may fail to capture objects at $45^\circ$ directions.}. $Area_{g_i}$ is the number of green pixels in the image for $i$th direction, and $Area_{t_i}$ is the number of total pixels in the image for $i$th direction. The vertical view angle is fixed to $0^\circ$, which is parallel to the horizontal line.

Green pixels are identified based on the RGB color band. A major limitation of this approach is the potential confusion of natural and artificial green objects: trucks painted green might be classified as green pixels, for instance. Some solutions have been proposed, including the use of image recognition technology such as semantic segmentation (\citet{cai2018treepedia}) and deep convolutional neural networks (\citet{cai2019quantifying}). However, as the precision of object recognition lies beyond the scope of this study, we consider the implementation of such advanced techniques future work.

Using metadata collected from the Google Street View API, it is  possible to know the months in which the images were taken. Since our interest lies in urban greenery, we defined "green months" for our study area as the period from April to October. Images taken outside of the "green months" are not utilized for analysis.

In addition to limiting images to green months, we implemented the functionality of specifying year of image via the Google Maps JavaScript API (\citet{li2016environmental}). This allows us to use images of the designated year when available, thus mitigating the bias of temporal fluctuation.

The code used in this study to calculate GVI is available on \href{https://github.com/y26805/Treepedia_Public}{GitHub}.

\subsection{Standardized Green View Index}
GVI is a site-based metric: it measures the visibility of surrounding greenery at a geographical point. However, in practice it is often aggregated to an area level (block, census tract, administrative boundary, etc.), in order to associate the index with other socioeconomic factors.
Previous studies (cf. \citet{li2016environmental}, \citet{lu2018effect}) implicitly assumed that each site of GVI calculation has equal importance, hence simply taking the mean of GVI scores of sites located in a given area. However, this assumption may not hold, especially if sites are not evenly distributed in space. A heterogeneous distribution of GVI points will lead to a biased estimation when aggregated to an area level, as points densely located will contribute to the aggregated value to a greater extent.

\begin{figure}[h]
\centering
\includegraphics[scale=0.4]{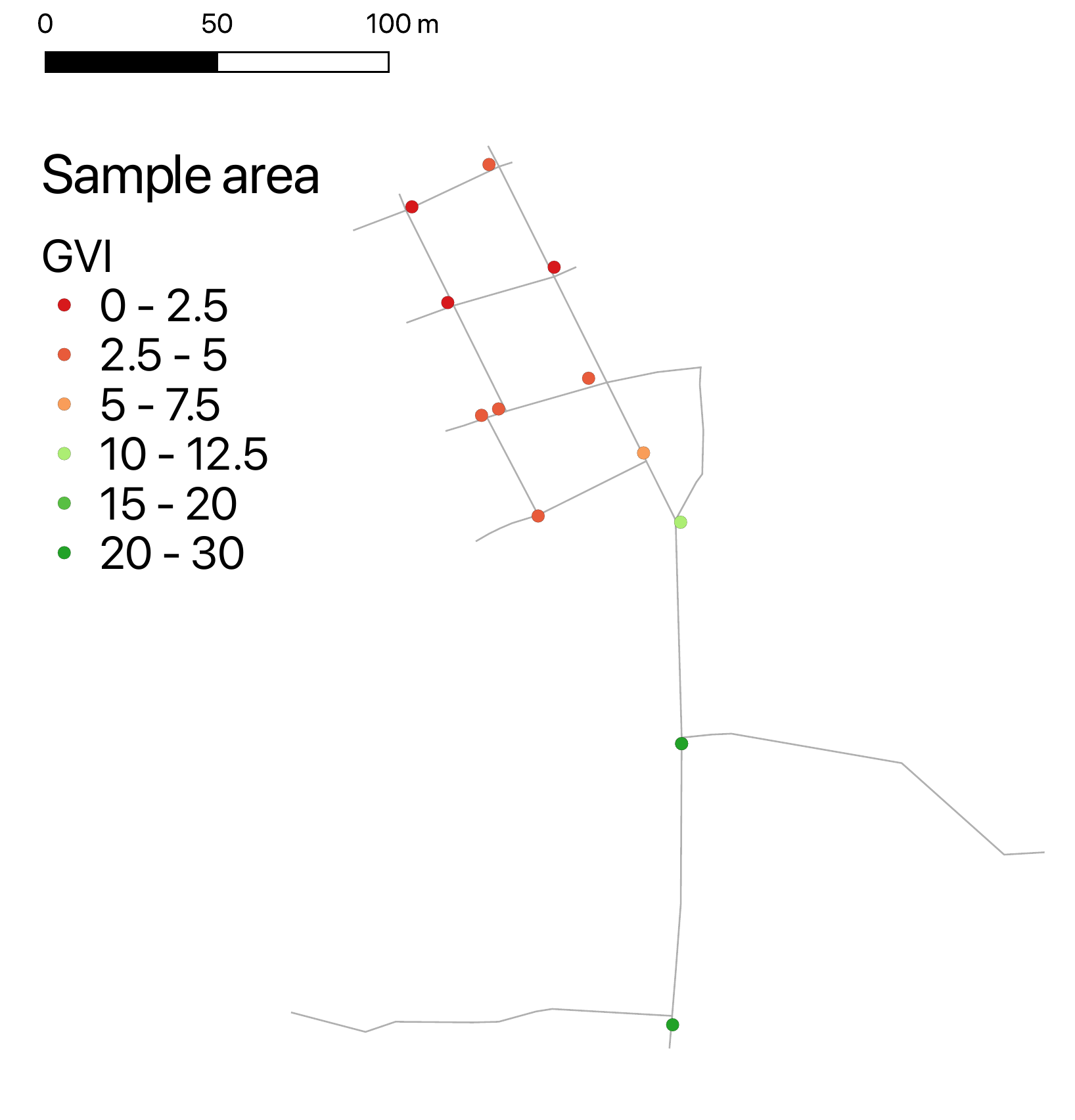}
\caption{Example of biased estimation of area-based GVI}
\label{img:validation}
\end{figure}

Figure \ref{img:validation} illustrates an example of such biased estimation. Sites with low GVI are densely located in the upper part of the area, while sites with high GVI are sparsely located in the lower part. Taking the mean of these sites skews the GVI value towards that of the denser parts (upper part of figure), resulting in a biased aggregation at the area-level. 

\begin{figure}[h]
\centering
\includegraphics[scale=0.51]{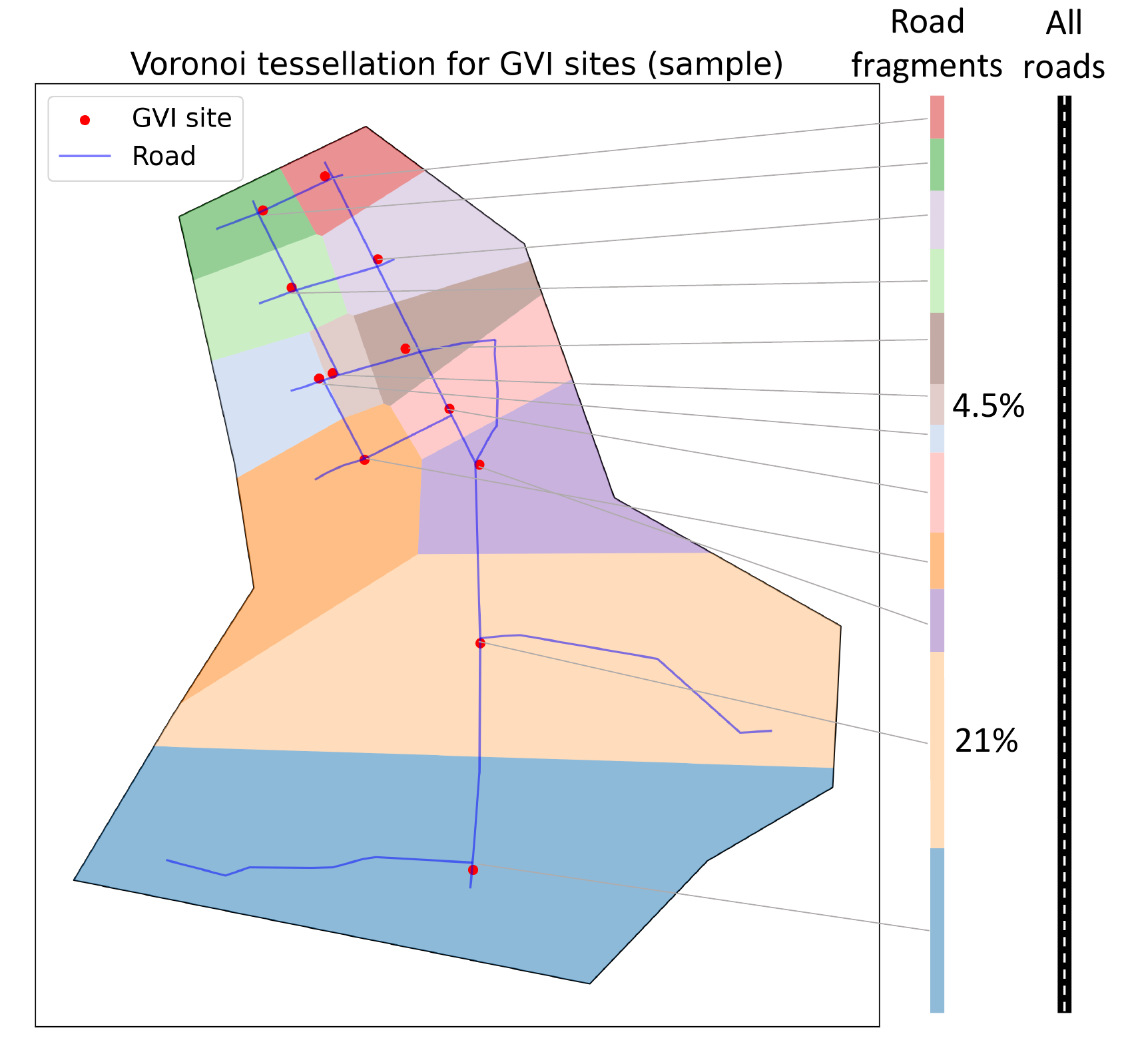}
\caption{Example of Voronoi tessellation and association of links.}
\label{img:voronoi}
\end{figure}

\newpage
In order to mitigate bias resulting from point density, we propose a Standardized Green View Index (sGVI) to calculate GVI for a given area. The sGVI is a weighted aggregation of GVI scores in a study area. It considers how road segments are located in the area when calculating the area-level value. The idea is to define the expected value of GVI in terms of total road length inside the zone to represent GVI of an area. In other words, sGVI is the expected value of GVI when the site for calculation is randomly chosen on the road network of the area. The mathematical formulation of sGVI is as follows:

\begin{equation} \label{eq:sgvi}
    sGVI = \sum_j GVI_j \times \frac{l_j}{l}
\end{equation}
where $j$ is point of GVI calculation $l_j$ is the total length of links that the point $j$ is associated with, and $l$ is the total length of all links in the zone. The association of point $j$ to links ($l_j$) is defined by the Voronoi tessellation of each point: the set of link fragments overlapped by the Voronoi tessellation is associated with point $j$. The procedure of this association is illustrated in Figure \ref{img:voronoi}.

Given a set of sites and an area, the Voronoi tessellation divides the area into cells so that any point in the area belongs to the cell of its closest site\footnote{This procedure is implemented using \href{https://github.com/WZBSocialScienceCenter/geovoronoi}{geovoronoi package} of Python.}. Once the tessellation is defined, the roads are overlapped and cut by each cell. Then the proportion of the road fragments contained in a cell over all the road fragments in the area is calculated, in terms of length of road fragment. These proportions are weighted inversely to the density of sites, which derives from densely located road segments.

One of the advantages of using the Voronoi tessellation over other methods is that every part of the roads is automatically associated with one site, without duplication. Given that the sites for GVI evaluation is located on a road segment, by definition, all the cells in the tessellation are supposed to have at least a fragment of road segment. In addition, since the cells do not overlay each other, it is ensured that one fragment of road segment is associated with only one site. This property may not be the case with other methods based on network configuration: for instance, if one decides to associate a site with nearby road segments, it is possible that two points in different road segments have the same distance from the site. An additional rule of attribution is needed in order to avoid duplicated association. The Voronoi tessellation, in turn, requires only the sites, the road segments and the boundary, and no other parameter is needed. As the equation \ref{eq:sgvi} indicates, quasi inverse weighting of GVI sites in terms of site density can be achieved with possibility of simple application in practice.

\subsection{Comparison of sGVI and other green metrics}
In order to explore the relation between sGVI and other green metrics, the following metrics were calculated in the study area of Yokohama city (see the next section for more detailed description): (1) sGVI, (2) GVI (mean), (3) GVI (median), and (4) NDVI. (1) sGVI is the expected value of GVI in the area, when a site is randomly chosen on the road network of the area. (2) GVI (mean) is an aggregation of GVI in the area by mean (\citet{lu2018effect}, \citet{li2018associations}), and (3) GVI (median) is an aggregation by median (\citet{li2016environmental}). (4) NDVI is an aggregation of NDVI in the area by mean\footnote{Note that NDVI contains some areas where GVI sites will not be located, since the GVI site is located on street. This may lead to an unfair comparison}.

For comparison, correlation among these four indicators is calculated. Then, the pair with the lowest correlation coefficient is further examined in terms of spatial distribution and regression analysis. The calculation of the green metrics was computed on an Intel Core i9 CPU at 2.4GHz and implemented in Python.

\subsection{Normalized Differential Vegetation Index}
While the GVI measures visible green vegetation on eye level, the Normalized Differential Vegetation Index (NDVI) quantifies the top-down green coverage using satellite imagery. The NDVI makes use of the fact that green vegetation reflects near-infrared lights more than red lights in the visible spectrum. The formula of NDVI is as follows:
\begin{equation}
    NDVI = \frac{NIR - Red}{NIR + Red}
\end{equation}
where $NIR$ is near-infrared light and $Red$ is red light. The value is normalized to $[-1,1]$, with a larger value signifying more abundant green vegetation.

The images used to calculate NDVI were retrieved from Level-2A data of Sentinel-2 with 10m resolution via \href{https://scihub.copernicus.eu}{Copernicus Open Access Hub}. Four images were retrieved for the study period\footnote{The dates were 9 March, 8 May, 5 August and 6 October, 2019.}, and in order to mitigate seasonal effects, mean values of each image for each pixel were taken. The cloud cover ratio of these images was smaller than 5\%.

\section{Case study: Yokohama city}
\label{section:4}
\subsection{Study area}
The green vegetation metrics defined in the previous section are applied in two wards (Nishi ward and Kanagawa ward) of Yokohama city, Japan. Located South of Tokyo metropolitan region, Yokohama city has its central business district located in the Nishi ward, and peripheral residential areas located in the West part of Kanagawa ward. This city structure allows us to study different behavior of the metrics, depending on the land use pattern. The geographical scale of analysis in this study is at Chome level (the Japanese name for a city block (\citet{gao2007effect})), since this is the smallest level of division for statistical data in Japan. 

\begin{figure}[h]
\centering
\includegraphics[scale=0.3]{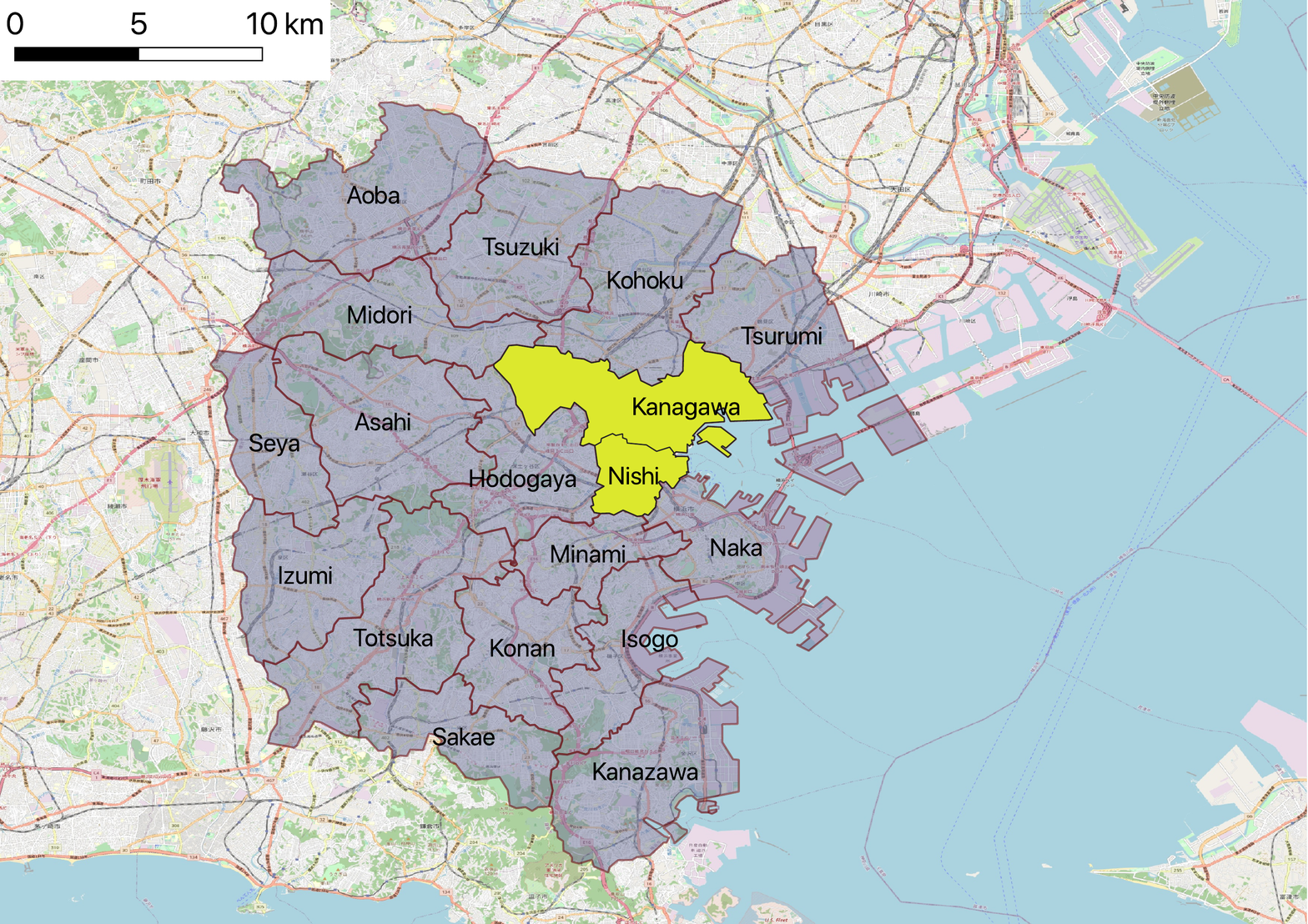}
\caption{Study area: Nishi ward and Kanagawa ward}
\label{img:study_area}
\end{figure}

The area and population of the study area is shown in Table \ref{table:yokohama} in Appendix \ref{ap:basic}. The locations of the study area is illustrated in Figure \ref{img:study_area} (\citet{yokohama2020}).

\newpage
\subsection{Descriptive analysis}
This section describes basic statistics of the data source that is used, namely GVI from Google Street View imagery and NDVI from satellite imagery.

\begin{figure}[h]
\centering
\includegraphics[scale=0.45]{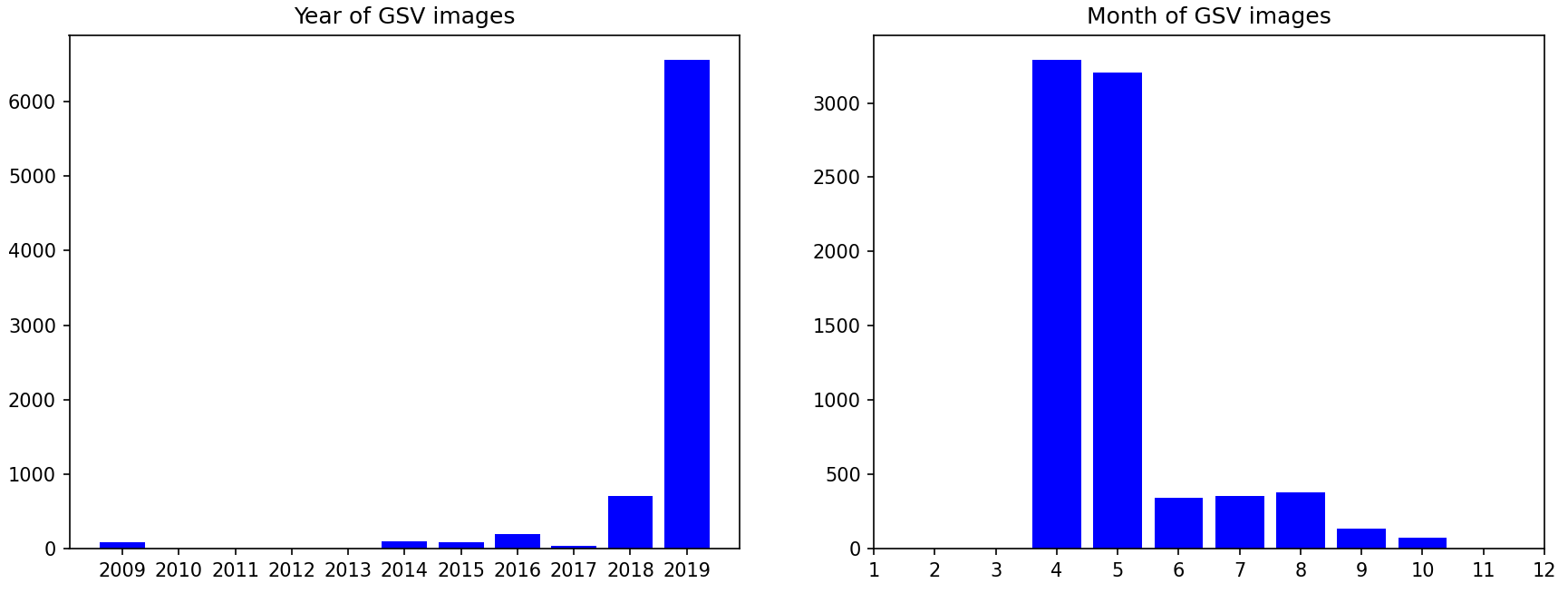}
\caption{Year and month of GSV images}
\label{img:gvi_year}
\end{figure}

For GVI, 7780 sites are selected in total, and  six corresponding images per site are retrieved via the Google Street View API. The sites are located every 100m along a link in the road network, and it is ensured that intersections have at least one site (if images satisfying criteria of season and year are available). 84\% of the sites turned out to have  images taken in 2019 (see Figure \ref{img:gvi_year}). This mitigates the time fluctuation, which has been pointed out in previous studies (\citet{Li2015a}, \citet{ye2019measuring}). Also, the month of image taking is limited to the period from April to October, with the majority (83\%) of images being taken in April or May.

As for NDVI, two satellite images are retrieved and the mean value for each mesh was calculated\footnote{Note that this aggregation will not produce spatial bias, because every satellite image has the same spatial resolution.}. The study area has in total 307,335 (70,273 (Nishi) + 237,062 (Kanagawa)) meshes with 10m spatial resolution. Figure 
\ref{img:ndvi} in Appendix \ref{ap:basic} illustrates the geographical distribution of NDVI.

\begin{figure}[ht]
\centering
\includegraphics[scale=0.5]{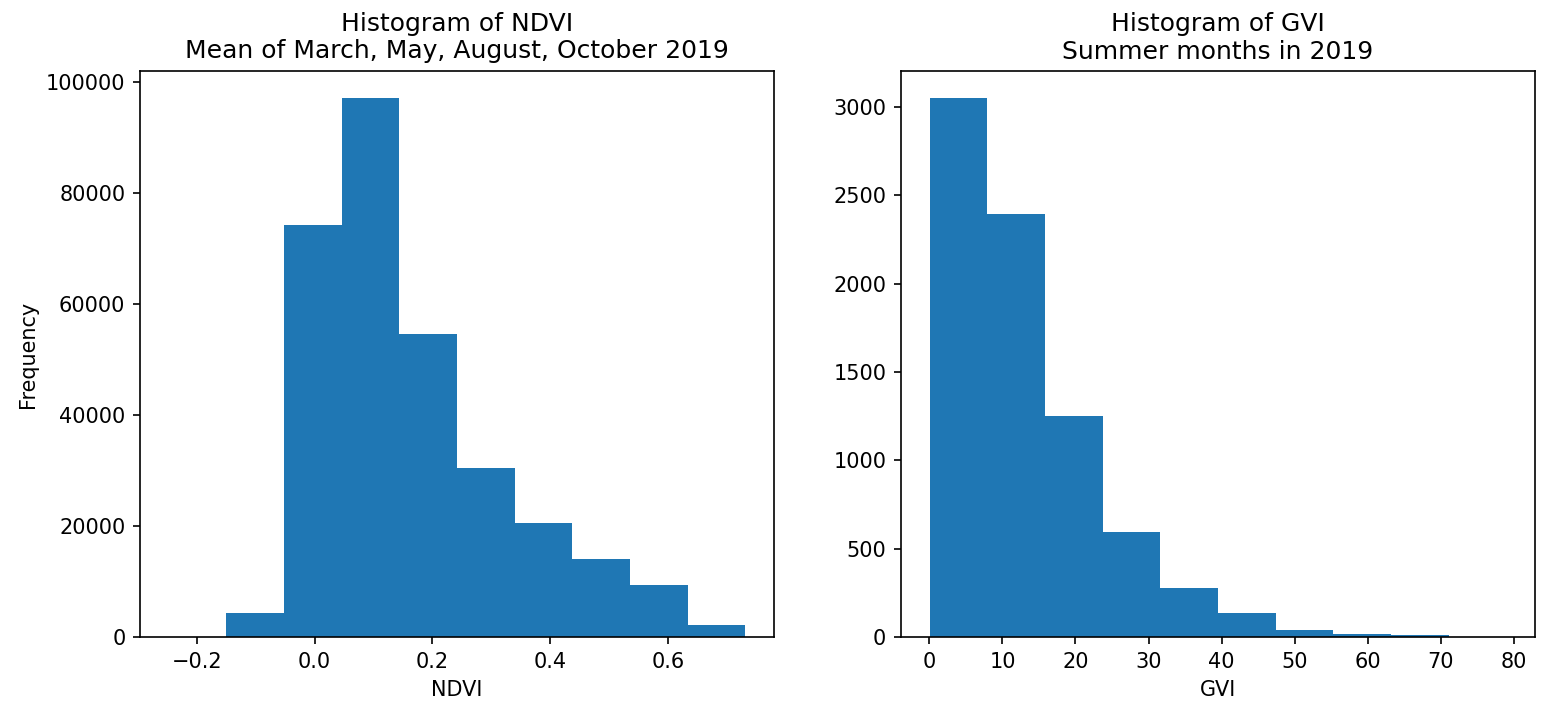}
\caption{Histogram of NDVI and GVI in the study area}
\label{img:ndvi_gvi_hist}
\end{figure}

Figure \ref{img:ndvi_gvi_hist} shows the histograms of NDVI and GVI for the study area, and the descriptive information is shown in Table \ref{table:green}.

\begin{table}[h]
\centering
\begin{tabular}{ccccccc}
Variable & N    & Mean & Median & Sd.  & min.  & max. \\ \hline
GVI      & 7,780 & 13.0    & 10.0   & 10.4 & 0.045 & 79.0 \\
NDVI     &  307,335   &   0.16    &  0.12 & 0.16  &   -0.25   & 0.73   \\ \hline
sGVI     & 213  & 10.2   & 10.0  &  5.7  & 0.0  & 32.0 \\
GVI(mean)   & 213 & 10.3 & 10.1 & 5.7 & 0.0 & 30.6 \\
GVI(median) & 213 & 9.4 & 8.7 & 5.6 & 0.0 & 24.9 \\
NDVI & 213 & 0.12 & 0.11 & 0.08 & -0.12 & 0.37 \\
\end{tabular}
\caption{Descriptive information of green metrics. The upper two variables are individual values (site and mesh), and the lower four variables are aggregated values at Chome level.}
\label{table:green}
\end{table}

\newpage
\subsection{Standardized Green View Index correcting spatial bias}

\paragraph{Example of biased aggregation}

Figure \ref{img:validation} showed an example of biased estimation of area-level aggregation of GVI resulting from heterogeneous site density. In the case of Figure \ref{img:validation}, the mean and median values of GVI are 8.38 and 3.53 respectively. If a researcher intends to study the relationship between these green metrics and socioeconomic factors in the area, it is evident that the choice of metrics will largely affect results of the analysis.

Admittedly, this situation does not occur in every study area, as the difference between mean and median values of GVI depends on the (numerical) distribution of GVI. For instance, if the GVI values in an area follow the normal distribution, the choice between mean and median will be unimportant since the two values are expected to be similar by definition. However, when the distribution is more long-tailed such as power low, the difference between mean and median is not always negligible, and the importance of defining a representative value will increase.

Even if the researcher has complete knowledge of the GVI distribution in the study area (which is divided into sub-areas), it is possible that the distributions that each sub-area follows are not identical. From this viewpoint, a feasible solution to mitigate this bias is to consider the weight of each site depending on their representativeness in the sub-area, which can be realized by sGVI.

\paragraph{Bias correction by sGVI}
The sGVI of the area shown in Figure \ref{img:validation} was 11.4, as sGVI places a greater emphasis on sites with high GVI in the lower part. In other words, when an arbitrary point on the network is chosen, the expected value of GVI there is 11.4. This result is also more intuitive, since the vegetation in the lower part will affect more the evaluation, covering larger superficial area.

The indicator is expected to behave as a proxy of representative value of areas, not only in areas with heterogeneous distribution of sites but also in areas with homogeneous distribution of sites. This is because, with such homogeneously located sites, sGVI leverages each point quasi-equally. If the distribution of GVI of such are is following normal distribution, the estimation by sGVI will be close to the mean and the median of GVI in the area. Such robustness towards distribution of sites and eventual variation of GVI is an advantage of sGVI, which has not been explored in previous studies.

\subsection{Comparison of different green metrics}
In order to further understand characteristics of the suggested indicator, sGVI, we first analyze the statistical relation between sGVI and NDVI. In this analysis, the Chome zones where sGVI is 0 are excluded. This is because of either lack of available image on Google Street View or absence of streets. With this pre-processed data set, Spearman's rank correlation\footnote{Pearson's correlation is not appropriate here because the distributions of sGVI and NDVI do not follow normal distribution, which is a prerequisite for performing Pearson's correlation test.} between the two metrics was calculated (Table \ref{table:cor}). The result was 0.72 (p$\ll$0.01). 

\begin{table}[h]
\centering
\begin{tabular}{ccccc}
            & sGVI  & GVI(mean) & GVI(median) & NDVI  \\ \hline
sGVI        & 1.000 & 0.953     & 0.926       & 0.727 \\
GVI(mean)   & 0.953 & 1.000     & 0.947       & 0.751 \\
GVI(median) & 0.926 & 0.947     & 1.000       & 0.678 \\
NDVI        & 0.727 & 0.751     & 0.678       & 1.000
\end{tabular}
\caption{Correlation matrix of green metrics}
\label{table:cor}
\end{table}

\begin{figure}[h]
\centering
\includegraphics[scale=0.5]{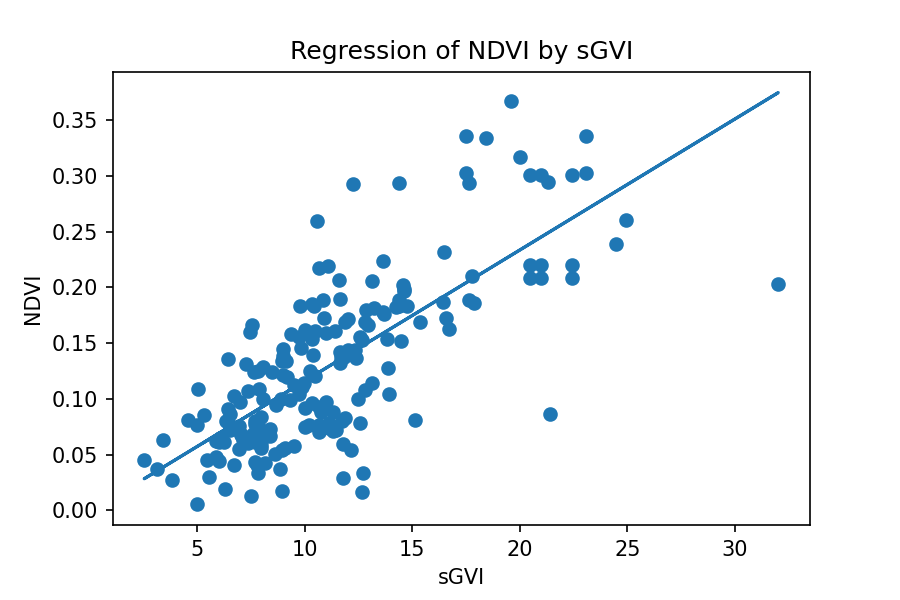}
\caption{Regression of NDVI by sGVI}
\label{img:reg_sGVI_NDVI_chome}
\end{figure}

Figure \ref{img:reg_sGVI_NDVI_chome} illustrates the scatter plot of sGVI and NDVI as well as the regressed line of NDVI by sGVI at Chome level, whose coefficient of determination was 0.57. From this result, we see that factors other than sGVI contribute to more than 40\% of the fluctuation of NDVI. Similar results were observed between GVI and NDVI in previous studies (\citet{larkin2019evaluating}, \citet{ye2019measuring}). This brings the question of what these other influencing factors are.

\begin{figure}[h]
\centering
\includegraphics[scale=0.42]{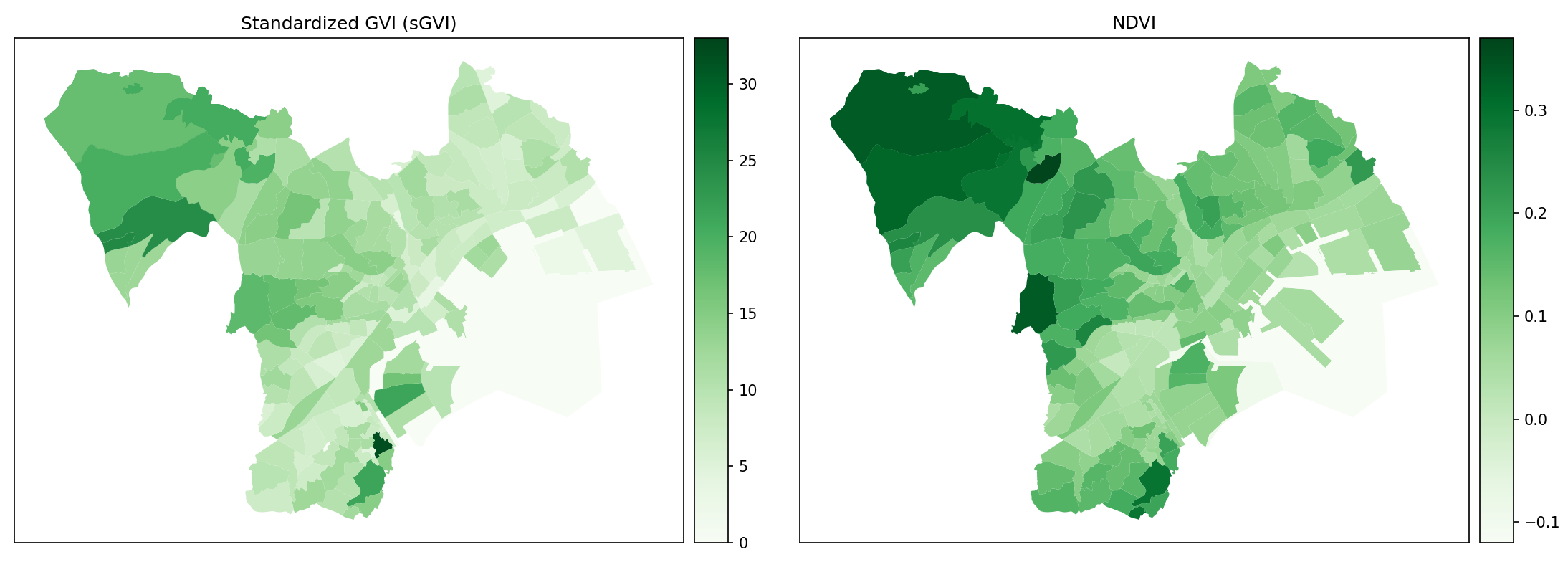}
\caption{sGVI and NDVI by Chome zone}
\label{img:sGVI_NDVI_chome}
\end{figure}

In the aim of exploring the above question, second, the geographical distribution of the estimated values was explored. Figure \ref{img:sGVI_NDVI_chome} visually compares sGVI and NDVI calculated at Chome level. Even though the scales of value between sGVI and NDVI are not directly comparable, there is an observable tendency that NDVI leverages the vegetation in the north-west part of the study area. 

\begin{figure}[h]
\centering
\includegraphics[scale=0.45]{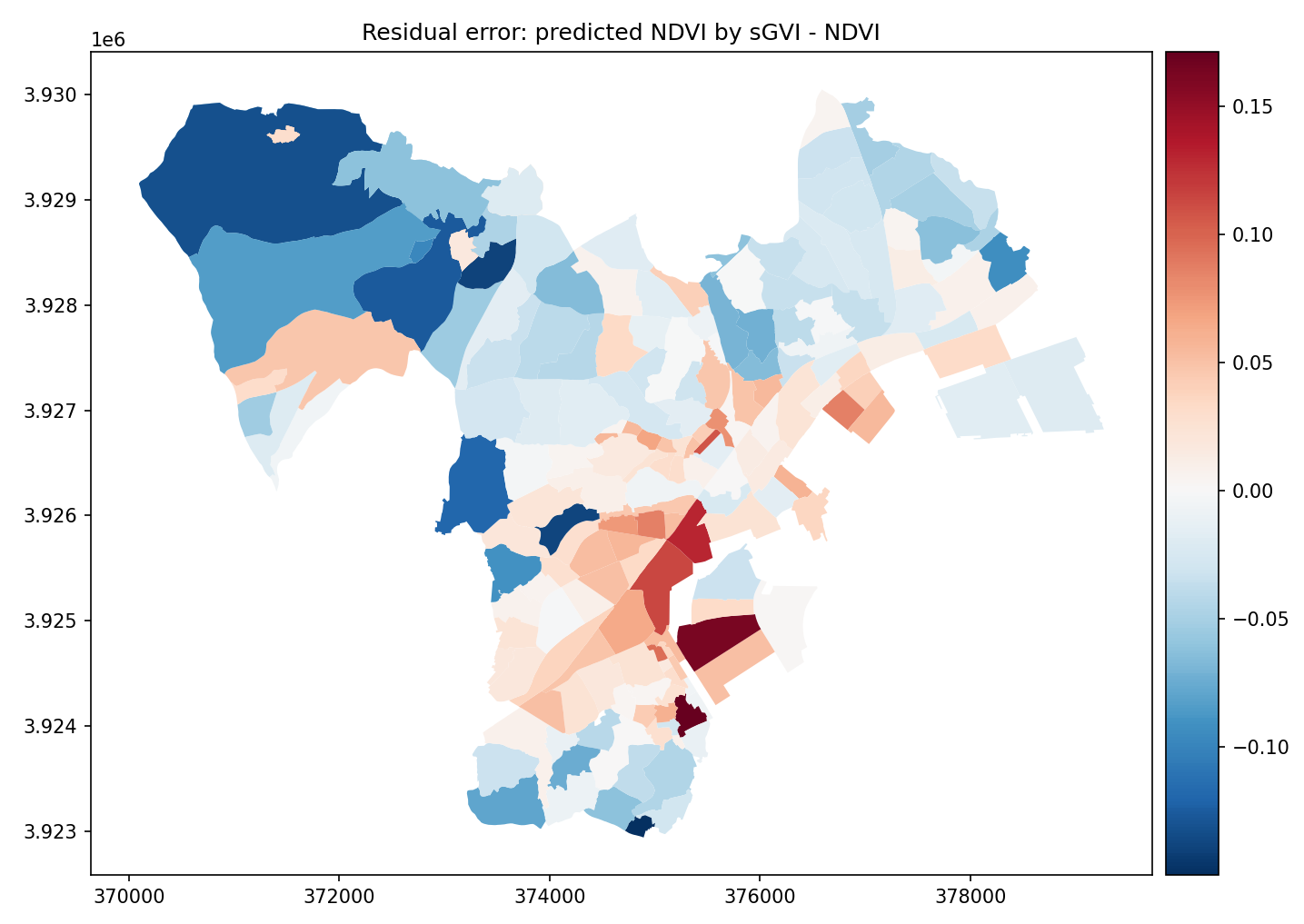}
\caption{Residual error of regressed NDVI by sGVI from sGVI}
\label{img:residual_sGVI_NDVI_chome}
\end{figure}

\newpage
Spatial distribution of residual error from the regression analysis is also illustrated in Figure \ref{img:residual_sGVI_NDVI_chome}. Under an assumption that the relation between sGVI and NDVI can be modeled by a linear function, NDVI in the north-west part where vegetation is more present is larger than what sGVI predicts, and NDVI in the south-east part where buildings are more dominant is smaller than the prediction by sGVI.

Given that the north-west part is mainly a residential area with parks and gardens, and that the south-east part is the central business district with limited number of vegetation along streets, this result indicates that sGVI captures the green vegetation in urbanized area more than NDVI. In other words, presence of buildings makes it difficult to estimate the amount of green vegetation from the top-down viewpoint.

\section{Discussion}
\label{section:5}
\subsection{Spatial aggregation of weighted points}
In geography, ``everything is related to everything else, but near things are more related than distant things” (\citet{tobler1970computer}). Spatial distribution of objects must therefore be considered when arguing at spatial level. From this viewpoint, simple aggregation of GVI values in a given area by mean signifies that every site is assumed to be related equally to every other sites, which is clearly false. If two sites are close, the images taken in the two sites are likely to capture an identical tree, whereas two distant sites do not have anything in common in their images. It is thus necessary to consider the spatial proximity of each site in order to estimate green vegetation, not for a site (point), but for a certain area.

This study proposed standardized GVI (sGVI), which is able to consider the heterogeneous relation among sites. Explicitly considering physical existence of road network, sGVI can mitigate a biased estimation caused by simply aggregating the mean value.

It is also possible to imagine other possibilities of weighting, especially using tools of spatial statistics. For instance, smoothing techniques such as Kernel Density Estimation (KDE) may be an option. However, KDE normally leverages densely located sites by definition, whereas the problem of aggregating GVI over an area is the opposite: provide less weight to densely located sites. This is why this method is not introduced in this study, but it may be a possible direction for future study.

\subsection{Different perspective: NDVI and GVI groups}
It has been pointed out that NDVI and GVI capture different aspects of green vegetation, due to their different viewpoints.

NDVI has a top-down viewpoint due to its use of satellite imagery, which makes it possible to capture horizontal extension of green vegetation better than other perspectives. Knowing that vegetation grows in a way that gives leaves maximum sunlight, it is reasonable to represent the existence of vegetation by NDVI. Nevertheless, it should be noted that there is green vegetation on building walls, especially in the city center, which is not fully captured by NDVI.

On the other hand, GVI has a street-level viewpoint. This is closer to humans' perception, but has several limitations. First, vegetation hindered by objects are not considered. Second, the estimation of GVI is limited on streets where pavement necessarily appears in the images, thus lowering the estimated value. Third, canopy formed by tall trees may be only partially perceived by people on the street, since the canopy will be placed at the marge of the eyesight, which is fixed in the horizontal direction. Lastly, densely located sites may be auto-correlated, because the same tree may be observed more than once, especially if the sites are very closely located.

These points themselves are consistent with the standpoint that GVI measures the visibility of urban green vegetation, but will lead inconsistency when GVI is employed as a proxy of the existence of vegetation. The usage of indicators deriving from GVI (including sGVI), thus, should be limited to measure visibility of green vegetation.

\subsection{Insight for further analytical work}
Multiple green vegetation metrics have been proposed, some of which are treated in this study. However, differences among them have not been fully understood. The result of this study, namely the comparison of NDVI and GVI, implies that the innate characteristic of each metric must be more carefully considered when associated with other factors.

Taking an example of physiological study, green vegetation is expected to have positive effects on physical health outcomes (\citet{larkin2019evaluating}). Nevertheless, when the causal relation between presence of green vegetation and health outcomes, there will be at least two paths: one is optical, and the other is atmospheric/olfactory/bacterial. The former corresponds to the effect of just ``seeing" greenery, whereas the latter corresponds to the effect of more direct interaction between human body and vegetation, such as quality of air, smell of plants and presence of certain bacteria. It is clear that GVI and FGVI (\citet{yu2016view}) is an appropriate method to measure the former, while NDVI is more suitable to evaluate the latter.

For statistical analysis, attention must also be paid to the aggregation method. Presence of several mediators between green vegetation and physiological outcome has been remarked (\citet{wang2019urban}), but spatial aggregation of site-based metrics has not been discussed. The proposed method of sGVI mitigates the bias from heterogeneous distribution of measurement points of GVI.

Association with these green vegetation metrics and other factors must be carefully discussed, knowing the properties of each method.

\section{Conclusion}
\label{section:6}
People's exposure to green vegetation is often associated with societal, economical or physiological factors, but it has been overlooked that aggregation method of green metrics may generate bias in area-based estimation. This study implemented GVI (\citet{Li2015a}) with development of designating time of image shooting, and proposed a new metric, standardized GVI (sGVI), which considers the density of sites for GVI calculation. We found that sGVI mitigate such density-led bias, and expect that it is more robust to heterogeneous spatial distribution of sites compared to simple aggregation by mean or median. Also, it was shown that NDVI leverages green vegetation in residential area with parks and gardens, while sGVI captures more vegetation in urban area where buildings are dominant. Therefore, for further analysis associating green vegetation and other factors, especially in urban areas, it is recommendable to employ sGVI since it mitigates bias from spatial distribution of sites and captures eye-level greenery in a more sensitive manner than NDVI.

For future works, there are two major issues to be considered. Firstly, the heavy computational load of Voronoi tessellation, especially when calculating sGVI in a large area, is not ideal. Keeping the idea of leveraging sparsely located sites, another direction with a lighter computational load must be explored.

The second issue is the treatment of missing points when estimating sGVI. Since not every site has images that satisfy the given conditions such as month and year, robustness to missing points must be considered in greater detail. While it is possible to mitigate this by placing sites with smaller intervals (20m instead of 100m, for example), it is not cost- efficient both in terms of time and money\footnote{Google Street View API is not free. Users are billed by the number of images requested via the API.}. 

Under SARS-CoV-2 epidemic, reduction of social contact by taking distance between people has been requested. Such situation may increase the importance of spaces outside buildings, and, thus, understanding the roles of the components in the outdoor space is crucial. From this view point, sGVI prepares a solid foundation on association studies, which will eventually contribute to the design of public spaces in the context of sustainable development.




\vspace{6pt} 

\appendix
\section{Miscellaneous information about the study area}
\label{ap:basic}
\begin{table}[h]
\centering
\begin{tabular}{cccc}
         & Area{[}$km^2${]} & Household & Population \\ \hline
Nishi    & 6.98          & 55 811    & 103 985    \\
Kanagawa & 23.59         & 126 093   & 245 036   
\end{tabular}
\caption{Statistics of the study area (2020.1.1)}
\label{table:yokohama}
\end{table}

\begin{figure}[h]
\centering
\includegraphics[scale=0.4]{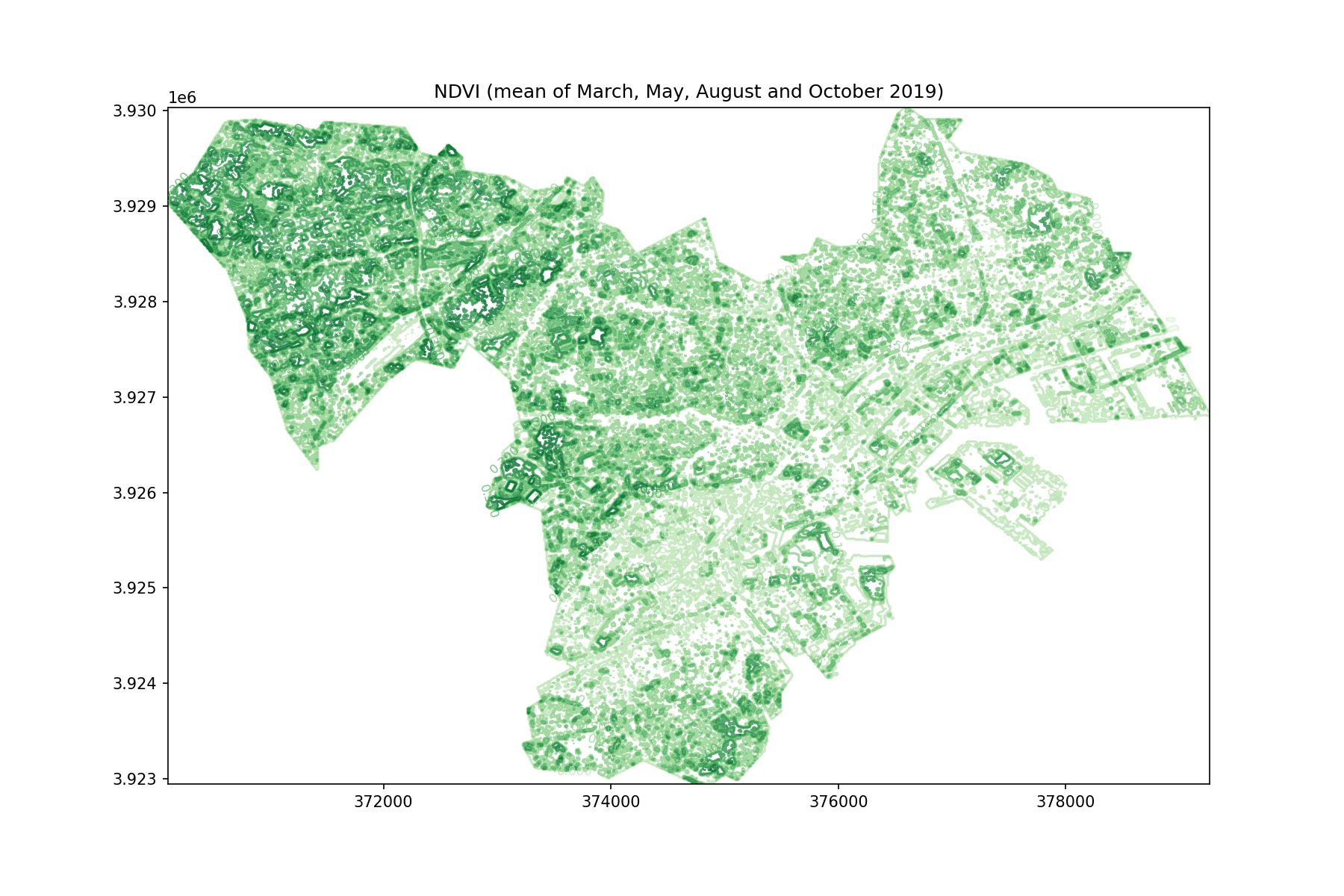}
\caption{NDVI in the study area}
\label{img:ndvi}
\end{figure}

\bibliographystyle{abbrvnat}  
\bibliography{references}  

\begin{thebibliography}{35}
\providecommand{\natexlab}[1]{#1}
\providecommand{\url}[1]{\texttt{#1}}
\expandafter\ifx\csname urlstyle\endcsname\relax
  \providecommand{\doi}[1]{doi: #1}\else
  \providecommand{\doi}{doi: \begingroup \urlstyle{rm}\Url}\fi

\bibitem[Astell-Burt and Feng(2019)]{astell2019association}
T.~Astell-Burt and X.~Feng.
\newblock Association of urban green space with mental health and general
  health among adults in australia.
\newblock \emph{JAMA network open}, 2\penalty0 (7):\penalty0 e198209--e198209,
  2019.

\bibitem[Boone et~al.(2009)Boone, Buckley, Grove, and Sister]{boone2009}
C.~G. Boone, G.~L. Buckley, J.~M. Grove, and C.~Sister.
\newblock Parks and people: An environmental justice inquiry in baltimore,
  maryland.
\newblock \emph{Annals of the Association of American Geographers}, 99\penalty0
  (4):\penalty0 767--787, 2009.

\bibitem[Cai et~al.(2019)Cai, Li, and Ratti]{cai2019quantifying}
B.~Cai, X.~Li, and C.~Ratti.
\newblock Quantifying urban canopy cover with deep convolutional neural
  networks.
\newblock \emph{arXiv preprint arXiv:1912.02109}, 2019.

\bibitem[Cai et~al.(2018)Cai, Li, Seiferling, and Ratti]{cai2018treepedia}
B.~Y. Cai, X.~Li, I.~Seiferling, and C.~Ratti.
\newblock Treepedia 2.0: applying deep learning for large-scale quantification
  of urban tree cover.
\newblock In \emph{2018 IEEE International Congress on Big Data (BigData
  Congress)}, pages 49--56. IEEE, 2018.

\bibitem[Chen et~al.(2019)Chen, Meng, Hu, Zhang, and Yang]{Chen2019}
X.~Chen, Q.~Meng, D.~Hu, L.~Zhang, and J.~Yang.
\newblock {Evaluating greenery around streets using baidu panoramic street view
  images and the panoramic green view index}.
\newblock \emph{Forests}, 10\penalty0 (12):\penalty0 1--14, 2019.
\newblock ISSN 19994907.
\newblock \doi{10.3390/F10121109}.

\bibitem[Gao and Asami(2007)]{gao2007effect}
X.~Gao and Y.~Asami.
\newblock Effect of urban landscapes on land prices in two japanese cities.
\newblock \emph{Landscape and urban planning}, 81\penalty0 (1-2):\penalty0
  155--166, 2007.

\bibitem[Gascon et~al.(2016)Gascon, Cirach, Mart{\'\i}nez, Dadvand,
  Valent{\'\i}n, Plas{\`e}ncia, and Nieuwenhuijsen]{gascon2016normalized}
M.~Gascon, M.~Cirach, D.~Mart{\'\i}nez, P.~Dadvand, A.~Valent{\'\i}n,
  A.~Plas{\`e}ncia, and M.~J. Nieuwenhuijsen.
\newblock Normalized difference vegetation index (ndvi) as a marker of
  surrounding greenness in epidemiological studies: The case of barcelona city.
\newblock \emph{Urban Forestry \& Urban Greening}, 19:\penalty0 88--94, 2016.

\bibitem[Gerrish and Watkins(2018)]{gerrish2018relationship}
E.~Gerrish and S.~L. Watkins.
\newblock The relationship between urban forests and income: A meta-analysis.
\newblock \emph{Landscape and Urban Planning}, 170:\penalty0 293--308, 2018.

\bibitem[Helbich et~al.(2019)Helbich, Yao, Liu, Zhang, Liu, and
  Wang]{helbich2019using}
M.~Helbich, Y.~Yao, Y.~Liu, J.~Zhang, P.~Liu, and R.~Wang.
\newblock Using deep learning to examine street view green and blue spaces and
  their associations with geriatric depression in beijing, china.
\newblock \emph{Environment international}, 126:\penalty0 107--117, 2019.

\bibitem[James et~al.(2015)James, Banay, Hart, and Laden]{james2015review}
P.~James, R.~F. Banay, J.~E. Hart, and F.~Laden.
\newblock A review of the health benefits of greenness.
\newblock \emph{Current epidemiology reports}, 2\penalty0 (2):\penalty0
  131--142, 2015.

\bibitem[Janh{\"a}ll(2015)]{janhall2015review}
S.~Janh{\"a}ll.
\newblock Review on urban vegetation and particle air pollution--deposition and
  dispersion.
\newblock \emph{Atmospheric environment}, 105:\penalty0 130--137, 2015.

\bibitem[Jesdale et~al.(2013)Jesdale, Morello-Frosch, and
  Cushing]{jesdale2013racial}
B.~M. Jesdale, R.~Morello-Frosch, and L.~Cushing.
\newblock The racial/ethnic distribution of heat risk--related land cover in
  relation to residential segregation.
\newblock \emph{Environmental health perspectives}, 121\penalty0 (7):\penalty0
  811--817, 2013.

\bibitem[Larkin and Hystad(2019)]{larkin2019evaluating}
A.~Larkin and P.~Hystad.
\newblock Evaluating street view exposure measures of visible green space for
  health research.
\newblock \emph{Journal of exposure science \& environmental epidemiology},
  29\penalty0 (4):\penalty0 447--456, 2019.

\bibitem[Lee and Maheswaran(2011)]{lee2011health}
A.~C. Lee and R.~Maheswaran.
\newblock The health benefits of urban green spaces: a review of the evidence.
\newblock \emph{Journal of public health}, 33\penalty0 (2):\penalty0 212--222,
  2011.

\bibitem[Li and Ghosh(2018)]{li2018associations}
X.~Li and D.~Ghosh.
\newblock Associations between body mass index and urban “green”
  streetscape in cleveland, ohio, usa.
\newblock \emph{International journal of environmental research and public
  health}, 15\penalty0 (10):\penalty0 2186, 2018.

\bibitem[Li et~al.(2015{\natexlab{a}})Li, Zhang, Li, Kuzovkina, and
  Weiner]{Li2015}
X.~Li, C.~Zhang, W.~Li, Y.~A. Kuzovkina, and D.~Weiner.
\newblock {Who lives in greener neighborhoods? The distribution of street
  greenery and its association with residents' socioeconomic conditions in
  Hartford, Connecticut, USA}.
\newblock \emph{Urban Forestry and Urban Greening}, 2015{\natexlab{a}}.
\newblock ISSN 16108167.

\bibitem[Li et~al.(2015{\natexlab{b}})Li, Zhang, Li, Ricard, Meng, and
  Zhang]{Li2015a}
X.~Li, C.~Zhang, W.~Li, R.~Ricard, Q.~Meng, and W.~Zhang.
\newblock {Assessing street-level urban greenery using Google Street View and a
  modified green view index}.
\newblock \emph{Urban Forestry and Urban Greening}, 2015{\natexlab{b}}.
\newblock ISSN 16108167.

\bibitem[Li et~al.(2016)Li, Zhang, Li, and Kuzovkina]{li2016environmental}
X.~Li, C.~Zhang, W.~Li, and Y.~A. Kuzovkina.
\newblock Environmental inequities in terms of different types of urban
  greenery in hartford, connecticut.
\newblock \emph{Urban Forestry \& Urban Greening}, 18:\penalty0 163--172, 2016.

\bibitem[Livesley et~al.(2016)Livesley, McPherson, and
  Calfapietra]{livesley2016urban}
S.~Livesley, E.~McPherson, and C.~Calfapietra.
\newblock The urban forest and ecosystem services: impacts on urban water,
  heat, and pollution cycles at the tree, street, and city scale.
\newblock \emph{Journal of environmental quality}, 45\penalty0 (1):\penalty0
  119--124, 2016.

\bibitem[Lu(2019)]{Lu2019}
Y.~Lu.
\newblock {Using Google Street View to investigate the association between
  street greenery and physical activity}.
\newblock \emph{Landscape and Urban Planning}, 2019.
\newblock ISSN 01692046.

\bibitem[Lu et~al.(2018)Lu, Sarkar, and Xiao]{lu2018effect}
Y.~Lu, C.~Sarkar, and Y.~Xiao.
\newblock The effect of street-level greenery on walking behavior: Evidence
  from hong kong.
\newblock \emph{Social Science \& Medicine}, 208:\penalty0 41--49, 2018.

\bibitem[Norton et~al.(2015)Norton, Coutts, Livesley, Harris, Hunter, and
  Williams]{norton2015planning}
B.~A. Norton, A.~M. Coutts, S.~J. Livesley, R.~J. Harris, A.~M. Hunter, and
  N.~S. Williams.
\newblock Planning for cooler cities: A framework to prioritise green
  infrastructure to mitigate high temperatures in urban landscapes.
\newblock \emph{Landscape and urban planning}, 134:\penalty0 127--138, 2015.

\bibitem[Nutsford et~al.(2013)Nutsford, Pearson, and
  Kingham]{nutsford2013ecological}
D.~Nutsford, A.~Pearson, and S.~Kingham.
\newblock An ecological study investigating the association between access to
  urban green space and mental health.
\newblock \emph{Public health}, 127\penalty0 (11):\penalty0 1005--1011, 2013.

\bibitem[Sanusi et~al.(2017)Sanusi, Johnstone, May, and
  Livesley]{sanusi2017microclimate}
R.~Sanusi, D.~Johnstone, P.~May, and S.~J. Livesley.
\newblock Microclimate benefits that different street tree species provide to
  sidewalk pedestrians relate to differences in plant area index.
\newblock \emph{Landscape and Urban Planning}, 157:\penalty0 502--511, 2017.

\bibitem[Tobler(1970)]{tobler1970computer}
W.~R. Tobler.
\newblock A computer movie simulating urban growth in the detroit region.
\newblock \emph{Economic geography}, 46\penalty0 (sup1):\penalty0 234--240,
  1970.

\bibitem[Tzoulas et~al.(2007)Tzoulas, Korpela, Venn, Yli-Pelkonen,
  Ka{\'z}mierczak, Niemela, and James]{tzoulas2007promoting}
K.~Tzoulas, K.~Korpela, S.~Venn, V.~Yli-Pelkonen, A.~Ka{\'z}mierczak,
  J.~Niemela, and P.~James.
\newblock Promoting ecosystem and human health in urban areas using green
  infrastructure: A literature review.
\newblock \emph{Landscape and urban planning}, 81\penalty0 (3):\penalty0
  167--178, 2007.

\bibitem[United{ }Nations(2018)]{un2018revision}
United{ }Nations.
\newblock World urbanization prospects - the 2018 revision.
\newblock Technical report, Department of Economic and Social Affairs, United
  Nations: New York, NY, USA, 2018.

\bibitem[Villeneuve et~al.(2018)Villeneuve, Ysseldyk, Root, Ambrose, DiMuzio,
  Kumar, Shehata, Xi, Seed, Li, et~al.]{villeneuve2018comparing}
P.~J. Villeneuve, R.~L. Ysseldyk, A.~Root, S.~Ambrose, J.~DiMuzio, N.~Kumar,
  M.~Shehata, M.~Xi, E.~Seed, X.~Li, et~al.
\newblock Comparing the normalized difference vegetation index with the google
  street view measure of vegetation to assess associations between greenness,
  walkability, recreational physical activity, and health in ottawa, canada.
\newblock \emph{International Journal of Environmental Research and Public
  Health}, 15\penalty0 (8):\penalty0 1719, 2018.

\bibitem[Wakefield et~al.(2001)Wakefield, Elliott, Cole, and
  Eyles]{wakefield2001environmental}
S.~E. Wakefield, S.~J. Elliott, D.~C. Cole, and J.~D. Eyles.
\newblock Environmental risk and (re) action: air quality, health, and civic
  involvement in an urban industrial neighbourhood.
\newblock \emph{Health \& place}, 7\penalty0 (3):\penalty0 163--177, 2001.

\bibitem[Wang et~al.(2019)Wang, Helbich, Yao, Zhang, Liu, Yuan, and
  Liu]{wang2019urban}
R.~Wang, M.~Helbich, Y.~Yao, J.~Zhang, P.~Liu, Y.~Yuan, and Y.~Liu.
\newblock Urban greenery and mental wellbeing in adults: Cross-sectional
  mediation analyses on multiple pathways across different greenery measures.
\newblock \emph{Environmental research}, 176:\penalty0 108535, 2019.

\bibitem[Wolch et~al.(2014)Wolch, Byrne, and Newell]{wolch2014urban}
J.~R. Wolch, J.~Byrne, and J.~P. Newell.
\newblock Urban green space, public health, and environmental justice: The
  challenge of making cities ‘just green enough’.
\newblock \emph{Landscape and urban planning}, 125:\penalty0 234--244, 2014.

\bibitem[Yang et~al.(2009)Yang, Zhao, Mcbride, and Gong]{Yang2009}
J.~Yang, L.~Zhao, J.~Mcbride, and P.~Gong.
\newblock {Can you see green? Assessing the visibility of urban forests in
  cities}.
\newblock \emph{Landscape and Urban Planning}, 91\penalty0 (2):\penalty0
  97--104, 2009.
\newblock ISSN 01692046.
\newblock \doi{10.1016/j.landurbplan.2008.12.004}.

\bibitem[Ye et~al.(2019)Ye, Richards, Lu, Song, Zhuang, Zeng, and
  Zhong]{ye2019measuring}
Y.~Ye, D.~Richards, Y.~Lu, X.~Song, Y.~Zhuang, W.~Zeng, and T.~Zhong.
\newblock Measuring daily accessed street greenery: A human-scale approach for
  informing better urban planning practices.
\newblock \emph{Landscape and Urban Planning}, 191:\penalty0 103434, 2019.

\bibitem[Yokohama{ }city(2020)]{yokohama2020}
Yokohama{ }city.
\newblock Yokohama city statistic gis data.
\newblock \url{
  https://www.city.yokohama.lg.jp/city-info/seisaku/torikumi/shien/gis/tokei-gis/gistat.html},
  2020.
\newblock visited: 2020-08-01.

\bibitem[Yu et~al.(2016)Yu, Yu, Song, Wu, Zhou, Huang, Wu, Zhao, and
  Mao]{yu2016view}
S.~Yu, B.~Yu, W.~Song, B.~Wu, J.~Zhou, Y.~Huang, J.~Wu, F.~Zhao, and W.~Mao.
\newblock View-based greenery: A three-dimensional assessment of city
  buildings’ green visibility using floor green view index.
\newblock \emph{Landscape and Urban Planning}, 152:\penalty0 13--26, 2016.

\end{thebibliography}






\end{document}